\documentclass[manuscript]{aastex}

\usepackage{graphicx}
\usepackage{dcolumn}
\usepackage{paralist}
\usepackage{comment}
\usepackage{graphicx,epsfig}
\usepackage{multirow}
\usepackage{wrapfig}
\usepackage{float}

\def\beq{\begin{equation}}
\def\eeq{\end{equation}}
\def\beqa{\begin{eqnarray}}
\def\eeqa{\end{eqnarray}}

\newcommand{\Nph}{N_{\rm ph}}

\def\erfc{\rm erfc}
\def\yAvg{\langle y \rangle}
\def\NF{{N_{\rm f}}}
\def\NFm1{{\NF{-}1}}
\def\ICSA{I_{\rm CSA}}
\def\as_p_s{{''\!/{\rm sec}}}
\def\d_p_d{{^\circ\!/{\rm day}}}

\slugcomment{To be sumitted to Ap J.}

\shorttitle{Detection of a faint fast moving asteriod}
\shortauthors{C.Zhai et al.}
\begin{document}


\title{Detection of a faint fast-moving near-Earth asteroid \\ using synthetic tracking technique}


\author{Chengxing Zhai,
Michael Shao,
 Bijan Nemati,
Thomas Werne,
Hanying Zhou,\\
Slava G. Turyshev,
Jagmit Sandhu
}
\affil{Jet Propulsion Laboratory, California Institute of Technology, \\
4800 Oak Grove Dr, Pasadena, CA 91109}
\email{chengxing.zhai@jpl.nasa.gov}
\and
\author{Gregg W. Hallinan,
Leon K. Harding
}
\affil{Department of Astronomy, California Insitute of Technology, \\
1200 E. California Blvd, Pasadena, CA, 91125} 

\begin{abstract}
We report a detection of a faint near-Earth asteroid (NEA), which was done using our synthetic tracking technique
and the CHIMERA instrument on the Palomar 200-inch telescope.
This asteroid, with apparent magnitude of 23, was moving at 5.97 degrees per day
and was detected at a signal-to-noise ratio (SNR) of 15 using 30 sec of data taken at a 16.7 Hz frame rate.
The detection was confirmed by a second observation one hour later at the same SNR.
The asteroid moved 7 arcseconds in sky over the 30 sec of integration time because of its high proper motion.
The synthetic tracking using 16.7 Hz frames avoided the trailing loss suffered by
conventional techniques relying on 30-sec exposure, which would degrade
the surface brightness of image on CCD to an approximate magnitude of 25.
This detection was a result of our 12-hour blind search
conducted on the Palomar 200-inch telescope over two nights on
September 11 and 12, 2013 scanning twice over six 5.0$^\circ\times$0.043$^\circ$ fields.
The fact that we detected only one NEA, is consistent with Harris's estimation
of the asteroid population distribution, which was used to predict the
detection of 1--2 asteroids of absolute magnitude H=28--31 per
night. The design of experiment, data analysis method,
and algorithms for estimating astrometry are presented.
We also demonstrate a milli-arcsecond astrometry using observations of two bright asteroids
with the same system on Apr 3, 2013. Strategies of scheduling observations
to detect small and fast-moving NEAs with the synthetic tracking technique are discussed.
\end{abstract}


\section{Introduction}
\label{sec:intro}

Recently, we introduced a technique of synthetic tracking that
enabled detection of small and fast moving near-Earth asteroids (NEAs) using short exposure frames
\citep{Shao2014}. Detecting and characterizing of small asteroids
is important for several reasons.
While being the subject of an interesting and
rapidly-evolving area of planetary science,
asteroids present a threat to the infrastructure and life on our planet.
For example, a 17~m asteroid that hit Russia on
Feb. 15, 2013, caused a severe damage to buildings and inflicted injuries to hundreds of people \citep{Brumfiel2013}. In addition, some NEAs may become targets for focussed 
space exploration efforts in the near future. In particular, characterizing small NEAs is needed to provide a potential target list for the upcoming NASA's asteroid redirection mission \citep{Lightfoot2013}. 

As discussed in \citep{Shao2014}, the synthetic tracking method processes the data from
a number of short exposure frames by shifting each frame according to a tracking velocity
vector so that the superposition of these shifted frames simulates a long-exposure integration
with the telescope tracking at that velocity.
This technique improves the detection's signal-to-noise ratio (SNR) by avoiding the trailing loss,
which typically affects the detection of fast-moving NEAs at distances $\lesssim$ 0.1 AU from the Earth.
This advantage is especially valuable for detecting small NEAs because these objects are observable
only at short distances. The improved SNR from using the synthetic tracking technique yields
a more precise astrometry of NEAs. In addition, synthetic tracking gains accruacy in astrometry
from the reduction of the effects due to atmospheric disturbances and imprecise telescope pointing
(to be addressed in Sec.~\ref{sec:astrometry}).

Synthetic tracking technique is made possible by the availability of new generation cameras
that provide both fast frame rate and low read noise.  For example,
scientific CMOS cameras can read at 100~Hz frame rate and only introduce read noise at 1e$^-$ level.\footnote{
\label{Andor2013} 
See a description of technical capabilities of the Andor's Neo and
Zyla sCMOS Cameras at: {\tt
http://www.andor.com/pdfs/literature/Andor\_sCMOS\_Brochure.pdf}}
Our observation on Palomar 200-inch used the Andor's
EMCCD$^{\ref{Andor2013}}$ operating at high EM gain of 200, making
the read noise benign compared with the sky background noise even
for the 16.7~Hz frame rate. This enabled us to detect a faint object
at the apparent magnitude of 23 using about 30 sec of data ($\sim$
500 frames). Without synthetic tracking, the detected asteroid with
speed 5.97 degrees per day ($\d_p_d$) moves about 7 arcsecond ($''$)
in the field. The corresponding surface brightness of the asteroid
yields approximately ${\rm SNR}\approx 4$ (apparent magnitude 24.5), below
the detection threshold of ${\rm SNR}=7$, set as our data processing
criterium. An additional benefit of synthetic tracking is
that it allows one to estimate velocity using only 30 sec of data,
making confirmation task much easier.

A 12-hour blind search was conducted on September 11 and 12, 2013, with six hours per night.
Our survey continuously scanned over the sky at a rate
of 5~$\as_p_s$, so that each star stays in the field for $\approx
30$~sec. It took about an hour to scan over each field of size
5.0$^\circ\times$0.043$^\circ$ (right ascention (RA) $\times$ declination (DEC)).
We repeated the scan in the next
hour to have two consecutive one-hour data covering the same field.
Thus, during each night we covered three different
5.0$^\circ\times$0.043$^\circ$ fields with a total of six fields.
the faint asteroid was detected in the second field on September 11,
2013, and confirmed by the repeated scan. Because this asteroid was
observed twice, both times with SNR of $\sim$15, the false positive
rate is practically zero. The asteroid moved 3770 pixels over 4626
seconds giving 5.97~$\d_p_d$ for the plate scale of 0.305~$''\!/$pixel.
Using estimation of the asteroid population  distribution from
\citep{Harris2011}, we expected to detect 1--2 asteroids of H
magnitude in range 28--31 per night. Our detection of one asteroid
in the two-night survey is consistent with this expectation.

The main step in the data processing with synthetic tracking technique is to search for signals
(or bright spots) in the synthetically integrated images for a grid of tracking velocities.
As mentioned in \citep{Shao2014}, this process is computationally intensive.
To overcome this difficulty, we use graphics processing units (GPUs) to perform synthetic tracking
at different velocities in parallel, thus, enable nearly real-time data processing.
To improve the SNR, we adopt a matched filter scheme \citep{Turin1960} to low-pass filter the data using the
point spread function (PSF) as the impulse response.
For convenience, we use a Moffat's PSF function template \citep{Moffat1969} as our PSF model
to quantify the shape of PSF and to reduce the computation in centroid fitting.
We use a bright star in the field to determined PSF model parameters.
A co-moving PSF fitting to the data frames is performed to generate precise astrometric solutions
and velocities for both asteroids and the stellar objects present in a frame.
The signal level resulting from the fitting is used to compute SNR and a false positive probability.
This approach yields the same SNR as the matched filter scheme
\citep{Gural2005, Shucker2008} in the case of using
a template filter velocity matching that of the asteroid.
The new feature in our approach is that we search in the tracking velocity space
for the faint object using parallel computing and then optimize the tracking veloicty
using the co-moving PSF fitting.

In addition to the improvement of the detection SNR, synthetic tracking also yeilds more accurate astrometry
for fast moving asteroids than the traditional long exposure approach.
We achieved milli-arcsecond (mas) level astrometry for two known bright asteroids,
observed on April 3, 2013, relative to nearby stars, after integrating over a minute using synthetic tracking.
However, if using long exposures, as simulated by co-adding the short exposure frames,
the astrometric precision does not improve after integrating
over 30 sec. This is because the effect due to atmosphere
and imprecise telescope pointing is no longer common between the asteroids and the background stars.
Working on short exposure images, synthetic tracking technique
makes the effect due to both the Earth's atmosphere and telescope pointing errors common
between the asteroids and the background stars and thus
achieves the similar precision of the relative astrometry
for the asteroids to that of the relative stellar astrometry\citep{Boss2009}.

This paper is organized as follows: In the Section~\ref{sec:syn_track}, we present the details of the synthetic tracking technique. The relevant algorithms are described in Section~\ref{sec:algorithm}. We present the data processing approach
in Section~\ref{sec:data_processing}. In Section~\ref{sec:results}, we describe the results of the detection of the faint asteroid and also demonstrate precise astrometry results using data from observing two known asteroids. Observation strategies using synthetic tracking for detecting small NEAs are discussed. We conclude in Section~\ref{sec:conclude}.

\section{Synthetic tracking}
\label{sec:syn_track}

\begin{figure}[ht]
\epsscale{0.6}
\plotone{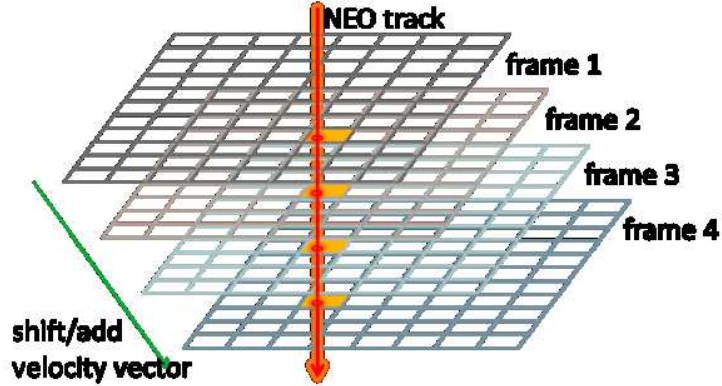}
\caption{Schematic chart showing how synthetic tracking aligns the data frames to track a asteroid by
displacing consecutive frames.\label{fig_synFrames}}
\end{figure}

As introduced by \citet{Shao2014}, synthetic tracking is a post-processing technique
that integrates a set of short exposure frames to simulate the tracking of the telescope
at a specific velocity, which we call the synthetic tracking velocity.
Figure \ref{fig_synFrames} shows the integration of the displaced frames according
to the velocity that is needed to track a NEA.
As this procedure is done in software, one has the flexibility of choosing all feasible tracking velocities.
For a NEA detection, we search in the synthetic tracking velocity space for a signal of the asteroid
without a trailing loss.
This is possible because a synthetically tracked image is very close to the image obtained with
a telescope that is actually tracking the asteroid.
With synthetic tracking both the moving asteroid and the background stars can be in sharp focus. 

Depending on the precision, at which the images are shifted,
synthetic tracking may be performed in two different ways,
which we call the Integer-Pixel-Shift-Add (IPSA) and
the Continuous-Shift-Add (CSA) methods respectively. We address both of them below.

\subsection{Integer-Pixel-Shift-Add synthetic tracking}
\label{sec:IPSA}

The IPSA synthetic tracking integrates frame images
by shifting them at an integer amount of pixels (i.e. no shift of image is done at a fraction of a pixel)
according to the tracking velocity
and then adding the signals collected on each of these frames. The quantity of interest here is the
integrated intensity of the synthetically tracked image,$I_{\rm IPSA}(x,y,v_x,v_y)$, which is computed as,
\beq
I_{\rm IPSA}(x,y,v_x,v_y) = \sum_{n=0}^{\NFm1}
I_n \Big(x + {\tt round}(n\, v_x), \; y + {\tt round}(n\,v_y)\Big) \,,
\eeq
where $n$ labels the frames and runs from $0$ to $\NFm1$.
The coordinates $(x,y)$ represent the location of a particular pixel on a CCD in (row, column) order,
with vector $(v_x, v_y)$ being the synthetic tracking velocity.
$I_n(x,y)$ is the intensity of the pixel at $(x,y)$ in the $n$-th frame.
Function ${\tt round}$ rounds its argument
to the nearest integer. Because the frames have zero bias (i.e.,
the background has been subtracted), in the case, when the round values are out of the bound of the frame index, zeros are filled in. To properly estimate the noise level,
it is necessary to record the actual number of frames that contribute to a signal
when its location is near frame boundary.
The frames that contribute ``filled zeros" to the signal should be excluded
because the signal is outside of their appropriate boundaries.
The advantage of the IPSA synthetic tracking is that it requires minimal number of arithmetic operations.
It is suitable for an extensive search for NEAs where many such operations are performed.
We implemented the IPSA synthetic tracking on GPUs to search for NEA signals over
different tracking velocities in parallel.

\subsection{Continuous-Shift-Add synthetic tracking}
\label{sec:CSA}

The CSA synthetic tracking method integrates short exposure frames
by shifting the frames according to the displacement determined by
the velocity of tracking, but not limited to integer number of pixels.
The images are displaced using the spectral interpolation or the Fourier space
interpolation method \citep{Boyd2011,Zhai2011}.
The spectral interpolation method for shifting the images is based on
the fact that the Fourier transforms of the original and shifted images
are related by a phase factor linear in spatial frequencies.
Mathematically, if we shift an image $I(x,y)$ by $(\Delta x, \Delta y)$ along row and column,
the displaced image is related to the original image via
\beq
  I^{\rm FT} \left (x + \Delta x, y + \Delta y \right )
= {\rm FT}^{-1} \Big\{ {\rm FT} \left \{
  I(x, y) \right \} e^{-2 \pi i \left ( k_x \Delta x + k_y \Delta y \right )} \Big\},
\label{spec_interp}
\eeq
where ${\rm FT}$ represents the Fourier transform
\beq
  {\rm FT}\left \{I(x,y)\right \} \equiv \sum_{x, y} I(x,y) e^{2\pi i \left ( k_x x + k_y y \right )},
\eeq
and ${\rm FT}^{-1}$ is the inverse Fourier transform.
 For a tracking velocity vector $(v_x, v_y)$, the CSA synthetic tracking image is then computed relying on the following expression
\beq
   \ICSA(x,y,v_x,v_y) = \sum_{n=0}^{\NFm1} I_n^{\rm FT} (x + nv_x, y + n v_y).
\eeq

\begin{figure}[ht]
\epsscale{0.75}
\plotone{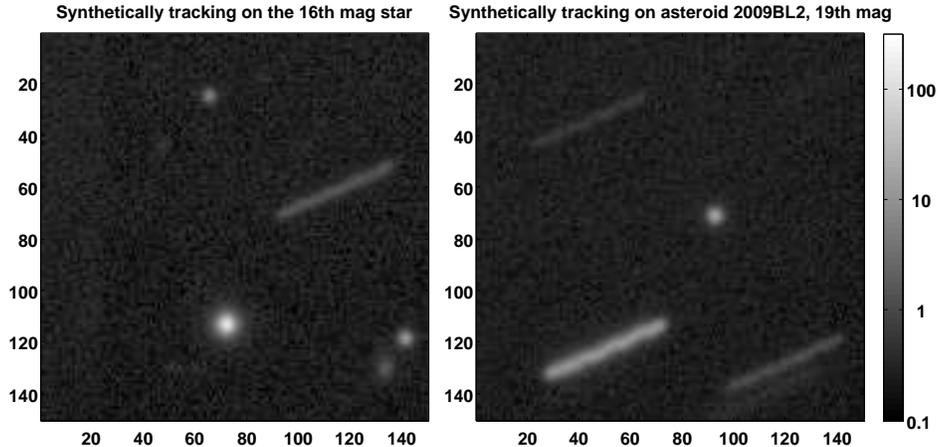}
\caption{Synthetic tracking images tracking on sky at the sidereal rate (left)
and tracking on asteroid 2009BL$_2$ (right), where stars are streaked.\label{fig_star_and_2009BL2}}
\end{figure}

To illustrate synthetic tracking technique, we use observational data  for asteroid 2009BL$_2$ collected on Palomar 200-inch on April 3, 2013. The short exposure frames were taken at a frame rate of 2~Hz. Figure \ref{fig_star_and_2009BL2} shows two synthetic tracking images for tracking the background stars (left) and the asteroid 2009BL$_2$ (right), respectively, while using CSA synthetic tracking with total of 960 frames of data.

\begin{figure}[ht]
\epsscale{0.55}
\plotone{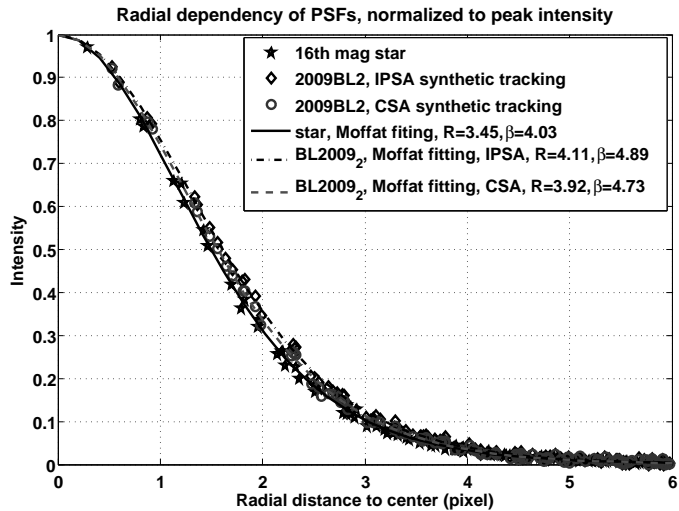}
\caption{Radial intensity profiles for the image of the 16th magnitude star
and synthetic tracking images of 2009BL$_2$ using IPSA and CSA respectively.
using IPSA and CSA respectively.\label{fig_star_nea_IPSA_CSA}}
\end{figure}

The bright background star has visual magnitude of 16 and the asteroid has apparent magnitude of 18.5. In the left image, the stars are tracked. The asteroid appears as a streak with surface brightness significantly less than that of the star. In the right image of Figure \ref{fig_star_and_2009BL2},
the asteroid is tracked and the stars are streaked. The surface brightness of the 16th magnitude star is now comparable with that of 2009BL$_2$, which is of 18.5th magnitude. The trailing loss as illustrated is approximately 2.3 visual magnitudes (a factor of 8.5). Due to the trailing loss, some of the faint background stars in left image
can barely be seen in the right image of Fig.~\ref{fig_star_and_2009BL2}. By synthetically tracking the stars and the asteroids, they have essentially the same PSF. To quantify the PSF, denoted as $P(x, y)$, we adopt the Moffat's PSF template \citep{Moffat1969}
\beq
  P(x, y) = \Big ( 1 + \frac{x^2 + y^2}{R^2} \Big)^{-\beta}
\label{MoffatPsf}
\eeq
to parameterize the measured PSF. Here, the quantity $R$ is a size scale and
$\beta$ specifies how fast the PSF falls off while moving away from the center.
The full-width-half-maximum (FWHM) of the Moffat's PSF is 
$W = 2R\sqrt{2^{1/\beta} - 1}.$
Fig.~\ref{fig_star_nea_IPSA_CSA} displays the radial
intensities of the synthetic tracking PSFs computed for the 16th magnitude star (star marker) and the asteroid 2009BL$_2$ (circle marker) respectively. The corresponding fitting curves using the Moffat's PSF are also displayed. The synthetic tracking PSFs of the star and asteroid are very similar, with  $W$ of the asteroid 2009BL$_2$ only 4\% larger than that of the star.

We note that even though the IPSA synthetic tracking only has resolution of one pixel, the achieved synthetic tracking image is very close to using the CSA synthetic tracking because the error due to neglecting sub-pixel (pix) displacement is much smaller than the size of the PSF ($\sim$3~pix). The third curve (diamond marker) in Figure~\ref{fig_star_nea_IPSA_CSA} shows the radial dependency of the intensity profile as function of the distance to the center of the image of asteroid 2009BL$_2$ using IPSA. The FWHM, is only 3\% larger for the IPSA synthetic tracking ($W=3.21$~pix) compared with the CSA synthetic tracking image ($W=3.11$~pix). The corresponding SNR is degraded by less than 3\%. While IPSA synthetic tracking is used by GPU search,
CSA is useful in post analysis for generating refined PSFs of tracked objects.

\section{Algorithms used in the processing}
\label{sec:algorithm}

In this Section we 
present the algorithm for estimating the background,
that was used extensively in our data pre-processing,
and the least-squares algorithm to perform fitting
for astrometric parameters, that was used in the post-analysis.
Application of both of these algorithms will be described in Sec.~\ref{sec:data_processing}.

\subsection{Estimating the background}
\label{sec_bg_est}

To detect faint objects, we need to accurately estimate the background bias to avoid too
many false positives (under-estimated background) or miss-detections (over-estimated background).
In the data frames, majority of the data points measure the sky background.
Only a small portion detects the light from the stars and asteroids.
Therefore, we can estimate the background by removing the signals, as outliers. This is done by starting with including all the data points in background data set and iteratively removing the signals that are above the average of background by a threshold, {\it e.g.}~5$\sigma$ ($\sigma$ is the standard deviation). 

The steps characterizing the background estimation procedure are given as follows: 
\begin{enumerate}[1)]
\item Let $y_i$ with $i\in\{1,2,\cdots, N\}$, represent all the data points available including both background and signal measurements. Let $B$ denote the set of the data points representing the background, which we need to determine. We initialize $B$ to be the entire set of data points and estimate $B$ iteratively.
\item Compute the sample mean $\yAvg$ and standard deviation of the data in set $B$, $\sigma_y$, as
\beq
  \yAvg = {1 \over N(B)} \sum_{i \in B} y_i \,,~~~~
  \sigma_y = \Big[{1 \over N(B)-1} \sum_{i \in B}\left (y - \yAvg \right )^2\Big]^\frac{1}{2},
\eeq
with $N(B)$ being the number of elements in the set $B$.
\item Update set $B$ according to the rule $B=\{ i | y_i \le \yAvg + \xi \sigma_y\}$, where $\xi$ is the threshold chosen for the estimation. 
\item Iterate this procedure by going back to Step 2) above until the process converges.
\end{enumerate}

If the population of the background is large compared with the signal data, this process converges fast,
because the background statistics, the mean and standard deviation, can be easily established and become stable.
In general, $\xi$ should be chosen according to the population of the sample. The larger $\xi$ is,
the faster the convergence can be reached. However, a larger threshold means a higher chance of including
weak signals into the background.
We used $\xi=5$ at the beginning to estimate an overall pixel independent background using 50 frames of data and
then used $\xi=4$ to estimate a pixel dependent background using about 2000 frames.

\subsection{Co-moving PSF fitting}

Co-moving PSF fitting optimizes the astrometry and velocity of the detected object by
a least-squares fitting of multiple short exposure data to a PSF that moves at the velocity of the object.
This assumes a {\it priori} knowledge of the PSF function.
It is performed after a synthetic velocity search, where the location and tracking velocity has been determined to certain accuracy depending on the velocity grid of the search. The estimations of the location and velocity
from synthetic velocity search are used as the initial condition for the optimization routines.
The mathematical formulation of the fitting is based on explaining the observed multiple
short exposure signals as a moving PSF, which can be expressed as
a minimization of the following least-squares cost function,
\begin{eqnarray}
  C\Big(v_x, v_y, x_c, y_c, \alpha, I_0\Big)&\equiv & \nonumber\\
&&
\hskip -130pt
\sum_{n=0}^{\NFm1}\sum_{x,y}
  \Big |I_n(x,y)
-\alpha P  \Big(
x-x_c-v_x \big (n-{\NFm1 \over 2}\big),  
y-y_c-v_y \big (n-{\NFm1 \over 2} \big)\Big ) {-}I_0 \Big | ^2 ,
\label{comoving_lsq}
\end{eqnarray}
where $P(x,y)$ is the PSF function, $(x_c, y_c)$ is the location
of the object at the mid epoch of all the frames, and $(v_x, v_y)$ is the velocity of the moving object (i.e., the telescope tracking velocity). Quantities $\alpha$ and $I_0$ are two extra fitting parameters specifying linear and constant levels with respect to the PSF.

We estimate the PSF $P(x,y)$ by fitting the Moffat's PSF template to a nearby bright star in the field. Because we critically sampled the PSFs \citep{Zhai2011}), in principle, we could reconstruct the PSF from the star image itself, which is especially important for micro-arcesecond
astrometry. For a milli-aresecond (mas) astrometry, it is sufficient to use the model for the PSF given by (\ref{MoffatPsf}) to
reduce the amount of numerical computation during centroiding fitting.

Minimizing the cost function $C(v_x, v_y, x_c, y_c, \alpha, I_0)$ gives an estimate
of the velocity of the asteroid $(v_x, v_y)$ and the location
of the object $(x_c, y_c)$ at the mid epoch of all the frames,
\beq
  \big (\hat v_x, \hat v_y, \hat x_c, \hat y_c, \hat \alpha \big ) = 
  \min_{v_x, v_y, x_c, y_c, \alpha, I_0} C \Big (v_x. v_y, x_c, y_c, \alpha, I_0 \Big ),
\eeq
where the estimate $\hat\alpha$ measures the signal level and is used to compute the SNR.
We adopted the Matlab {\tt lsqnonlin} routine to perform this optimization.
We note that the co-moving PSF fitting can be applied when the objects are so faint
that neither the star nor the asteroid are detectable in a single frame.
The proper convergence relies crucially on the {\it priori} knowledge of the location
and velocity of the detected faint object from the synthetic tracking search.
Using the estimated velocity $(\hat v_x, \hat v_y)$, we use the CSA synthetic tracking to
obtain a refined PSF of the tracked object.

With estimate $\hat\alpha$ in hand, we express the detection SNR after applying
the matched filter as
\beq
  {\rm SNR} = {\hat \alpha \sqrt{\NF}\sqrt{\sum_{x,y} P(x,y)^2} \over \sigma_n} \,,
\label{detection_snr}
\eeq
where the sum is over pixels and $\sigma_n$ is the standard deviation of
the noise detected by each pixel (assumed to be uniform across pixels).
Factor $\sqrt{\NF}$ is from integrating $\NF$ independent frames.
A convenient approximation that expresses the SNR in terms of $\Nph$ and the FWHM of the PSF is
\beq
  {\rm SNR} = 0.6 {\Nph \sqrt{\NF}  \over \sigma_n W} \,.
\eeq

\subsection{Centroiding and 
estimation of the tracking velocity}

\begin{figure}[t!]
\epsscale{0.65}
\plotone{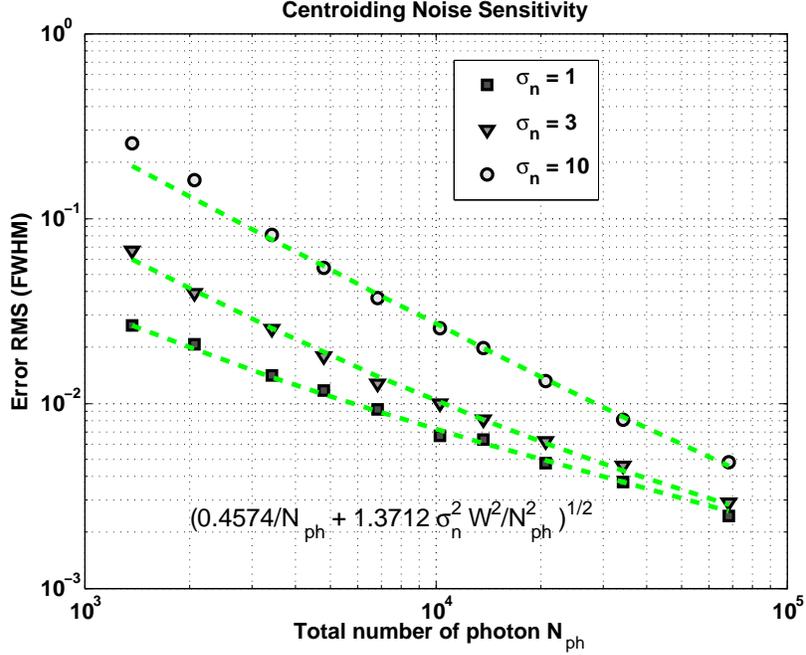}
\caption{The astrometry precision as function of the total number of photons for
three different background noise levels, specified by $\sigma_n$,
which is the standard deviation of noise for each pixel.\label{fig_snr}}
\end{figure}

We studied the sensitivity of astrometric solutions to the noise present in the system
using simulations.
For this, we adopt a Moffat's PSF with $R \approx 3.3, \beta \approx 4$, that were obtained from
fitting the 16th magnitude star images in the field taken while
observing 2009BL$_2$. The co-moving PSF fitting procedure is applied to simulated signals to
estimate the velocity and astrometric position of the moving object. 
The simulation was performed for many different signal levels and three different noise levels.
Figure~\ref{fig_snr} displays the astrometric error RMS as a function of the total number of photons
collected for three different levels of background noise, parameterized by $\sigma_n$,
the standard deviation of the background noise per pixel.  The squares, triangles, and circles
represent the three different noise levels, respectively. By inverting
the Hessian matrix \citep{Press1986} of the least-squares cost
function (\ref{comoving_lsq}) for a Moffat's PSF and assuming Poisson statistics
for photon detection, we derived the
following empirical formula to assess the uncertainty for astrometric
position
\beq
   \sigma_{x,y} = \sqrt{{0.457 \over \Nph} + {1.37 \sigma_n^2 W^2 \over \Nph^2}} {W \over \sqrt{\NF}},
\label{eq:sig}
\eeq
where $W$ is the FWHM of the PSF in 
pixels and $\Nph$ is the total number of the photons detected per frame. 
Fig.~\ref{fig_snr} shows that the estimated RMS using simulation agrees well with the empirical formula.
For faint objects, the background noise, given by the second term under
the square root in Eq.~(\ref{eq:sig}), dominates. In this case, it is convenient to approximate Eq.~(\ref{eq:sig}) as 
\beq
   \sigma_{x,y} \approx {0.65 W \over {\rm SNR}} \,,
\label{cent_snr}
\eeq
where ${\rm SNR}$ is the signal-to-noise ratio for detection given by Eq.~(\ref{detection_snr}).

Precision of the synthetic tracking velocity  is related to the centroiding precision via 
\beq
 \sigma_{v_x,v_y} = {\sqrt{12} \, \sigma_{x,y} \over T_{\rm tot}} 
\label{vel_snr}
\eeq
where $T_{\rm tot}$ is the total time duration covered by all the short exposure frames.

\section{Data processing method}
\label{sec:data_processing}

\subsection{Observation and data processing overview}
We conducted a 12-hour blind search using the Caltech HIgh-speed Multi-color camERA (CHIMERA)\footnote{\label{CHIMERA} Details for the Caltech HIgh-speed Multi-color camERA (CHIMERA) placed at prime focus of the Palomar 200-inch are availble at: http://www.tauceti.caltech.edu/chimera/}
at the Palomar 200-inch telescope over two nights on September
11-12, 2013. CHIMERA uses two Andor iXon3 888 EMCCDs with
1024$\times$1024  pixel detector allowing readout at 10 MHz data
rate in two colors with very low effective read noise of
$\ll 1{\rm e}^-$ with EM gain applied. We use this instrument to take images
at 16.7 Hz  with EM gain of 200 to avoid excessive read noise. This
allowed us to scan over the sky at 5~$\as_p_s$ rate continuously
instead of slewing and stopping the telescope repeatedly. Each scan
was along the RA direction and lasted for
approximately 1-hour to cover a field of size
5.0$^\circ\times$0.043$^\circ$. Each object was observed in the field of view (FOV) for about
30~sec. We binned 2$\times$2 CCD pixels to have 512$\times$512 frame
pixels. The FOV is approximately 2.6 arcminute (or 0.043$^\circ$).
Our observation was carried out without any additional
optical filter, which could be optionally applied. During each
night, we divided the allocated 6-hour observation into three
two-hour pairs, with each pair scanning over the same field twice,
to cover three different fields.  To facilitate data processing,
we took calibration data sets including very short exposure
(10~$\mu$sec) frames for estimating frame bias and flat field
responses using twilight. At a future time when we are able to
process the data in real time, the repeated one hour observation
would only be conducted when we have detected a moving object.

\begin{figure}[ht]
\epsscale{0.50}
\plotone{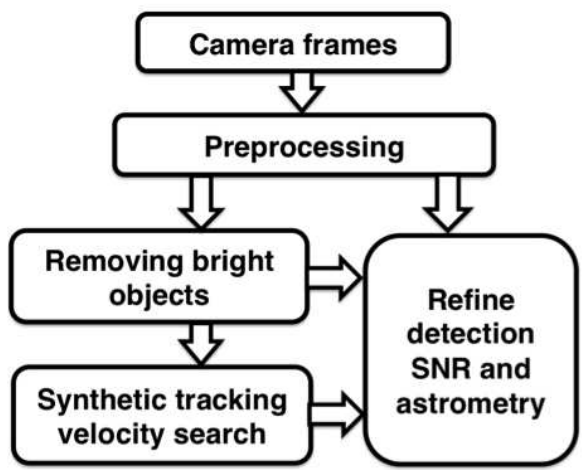}
\caption{A flow chart of data processing for detecting NEAs using synthetic tracking
on multiple short exposure frames. \label{fig_flowChart} }
\end{figure}

The entire data process contains four main steps:
preprocessing, detecting and removing objects from frames, synthetic tracking velocity search, post analysis to refine the results. Figure~\ref{fig_flowChart} displays a flow chart showing the relationship between them.

In pre-processing, we adjust data frames using calibration data of the A2D bias and flat field response.
A pixel-dependent sky background is estimated and subtracted to make
the data frames zero biased. We remove the bright pixel data caused by cosmic ray events by setting the values to zero. In the second step, we detect stationary objects (stars and galaxies) as well as bright asteriods and remove their signals from each frame by setting the relevant pixel values to zero. The data is then passed to the synthetic
tracking velocity search, where the bright signals above the detection threshold in the synthetically tracked image at each of the grid velocities are detected.
The results are then reported for the post analysis to refine the astrometry and detection SNR using optimization.

Because the PSF is larger than a CCD pixel, the matched filter technique \citep{Turin1960} is used
to improve the SNR. The matched filter convolves the data with a low pass filter, whose impulse response profile matches the PSF and yields the optimal SNR. We now provide details for each data processing step.

\subsection{The pre-processing procedure}

In pre-processing, we first deal with known features in the measurements.
Two camera calibration data sets were taken to measure the A2D bias frame
(corresponding to zero intensity) using very short exposure time of 10 $\mu$sec
and the flat field responses using twilight.
To enable asteroid detection the bias map is subtracted from each of the data images.
All the frames are then divided by the flat field responses to compensate for pixel dependent throughput.
Next we remove the cosmic ray events, which generate high counts that only stay in a single frame.
Cosmic ray events are detected by first zeroing out the bright stars that have signal level
comparable with the cosmic ray events in a single frame. (We differentiate the
bright stars from the cosmic ray events by integrating over hundreds of frames
so that they are much brighter than the extra counts due to cosmic ray events, which only appear in one frame.)
After zeroing the signals from the bright stars, differences between consecutive frames are used to identify highly
varying signals on the frame-to-frame basis and attribute them to cosmic ray events.

We then estimate a pixel-dependent background intensity and an average background noise level using about 2000 frames of data. To do this, we first estimate a uniform background (pixel independent) and noise level by computing the sample mean and standard deviation of background data
in a small number (e.g., 50) of frames. It is convenient
to divide all the data points into two parts, background data and signals, assuming signals are above the background for a specified threshold. The background data set is
estimated by iteratively removing signals above the sample mean by 5$\sigma$. (See Sec.~\ref{sec_bg_est} for the details of the algorithm.) Using this initial estimation of the background and noise levels, we are able to eliminate most of the strong signals from each frame. 

For the rest of the data, we apply the same algorithm to the data measured by each pixel (about 2000 frames) to estimate a pixel-dependent background using a threshold of 4$\sigma$.
After removing signals from each of the frame, we further detect and remove background stars that has SNR above 4 after integration over 30 sec (average observation time
for each object in the field for the scan rate we used) by performing a synthetic tracking at the sidereal rate. 
After removing bright objects and faint stars, the field-dependent background is estimated by
taking average over the background data (typically more than 1000 data after removing signals)
for each pixel. We also compute the frame-to-frame variation of the background for each pixel
and then take the average over all the pixels as an estimate for the noise level of background.
The estimated sky background is subtracted from each frame to have zero biased frames.
Note that for estimating background, we used a threshold that is much lower than our object
detection SNR threshold of ${\rm SNR}=7$.

\subsection{Detecting and Removing objects for synthetic tracking velocity search}

Before searching for faint asteroids, we need to detect and remove the bright objects from each frame. With the estimated background noise level, it is straight forward
to detect all the bright objects in each of the zero biased data frames with single frame SNR above threshold 7. We then synthetically track at the sidereal rate to detect faint static objects with SNR above 7 after 30 sec of integration. The detected objects are passed to post-analysis for further identification. For example, a very bright asteroid may be detected if its single frame signal is above the specified detection threshold. Also, if an asteroid does not move much during the observation period, it may be detected as a static object. We remove the signals of the detected objects by setting the values at the relevant pixels to zero. Before passing to the next step, we apply the matched filter by convolving each data frame with the PSF profile to lessen computational operations during the extensive synthetic tracking velocity search.

\subsection{Synthetic tracking velocity search using GPUs}

Synthetic tracking velocity search examines the data over a two
dimensional tracking velocity grid. To avoid trailing loss, the
grid spacing should be no more than the size of the PSF divided by
the integration time, i.e. the speed at which the motion is less
than the size of PSF over the integration time (30 sec in our case).
An IPSA synthetic tracking is performed for each grid velocity.
Because the frames have zero background, we fill in zeros for the
missing data at the boundaries of frames that are displaced. If the
synthetic tracking image shows a signal above the noise level by
the detection threshold of 7, we report this signal level together
with four numbers $(x,y,v_x, v_y)$, where $(x,y)$ specifies the
location of the detected signal and $(v_x, v_y)$ is the grid
velocity at which the synthetic tracking yields the signal. The
post-analysis uses these information to refine the detection SNR
and to compute astrometry using optimization schemes.
Because data processing for the synthetic tracking search  is independent between different tracking velocities, it can be easily speeded up by implementing parallel computing. We have implemented the search using the NVIDIA's Tesla K20c GPU\footnote{\label{NVIDIA} Information NVIDIA's Tesla GPU Accelerators may be found at {\tt http://www.nvidia.com/object/tesla-workstations.html}}
to accelerate the faint object search. The
K20c is based on the company's state-of-the-art Kepler architecture equipped with 2496 CUDA processing cores and 5~GB of GDDR5 RAM. Peak single-precision processing performance is 3.52~TFLOPS ($10^{12}$ floating operations per second). Our performance is currently limited by the memory bandwidth, which is 209~GB/sec. The search software is implemented in C/C++. To process a 90 sec data cube of 1500 frames of 512$\times$512 images searching over a 100$\times$100 synthetic tracking velocity grid covering a velocity range of $\pm$12~$\d_p_d$ in both RA and DEC, the average time of GPU is under 90 sec. Our current grid spacing is 1~pixel/integration time, which is finer than needed (PSF size is $\sim$3~pix). Using a coarser grid, this performance allows a real time processing. Our detection threshold is set to ${\rm SNR}=7$ to have less than 1\% false positive probability per 30 sec of data.

\subsection{Post-analysis, computation of detection SNR and astrometry}

With the information from synthetic tracking velocity search that the detected signal is above ${\rm SNR}=7$, we refine the detection using an optimization scheme by fitting the data frames to a co-moving PSF as described in the algorithm Sec.~\ref{sec:algorithm}. The least-squares fitting yields an optimal velocity for the moving object and the astrometry relative to the camera frames. We obtain solutions for stellar astrometry in the same fashion to compute the relative astrometry of the asteroid with respect to stars. The fitting results yield also the signal strength above the background, which is described by $\alpha P(x,y)$, with $\alpha$ determined by the fitting procedure. The detection SNR can be then computed using Eq.~(\ref{detection_snr}).

We estimate the false positive probability for initial detection as
\beq
  P_{\rm false-alarm}^{\rm detection} = N_x N_y N_{v_x} N_{v_y} \times {1 \over 2}
   {\tt erfc} \left ( {\rm SNR } \over \sqrt{2} \right),
\label{Prob_false}
\eeq
where ${\tt erfc}$ is the Gaussian complimentary error function, $N_x, N_y$
are the dimensions of the CCD, and $N_{v_x}, N_{v_y}$ are dimensions of the synthetic tracking velocity search grid along RA and DEC, respectively. This is derived by assuming the background fluctuates independently
with a Gaussian statistics so that the total false detection probability is
the number of trials, which is the product of total number of pixels and total number of grid velocitys
$N_x N_y N_{v_x} N_{v_y}$,
multiplying the false detection probability of one trial $\erfc ({\rm SNR}/\sqrt{2})/2$.
We choose ${\rm SNR}=7$ as the threshold for $N_x=N_y=512$, $N_{v_x}=N_{v_y}=100$, in order to have false alarm rate less than 1\% ($\sim$ 0.34\%) per 30 sec data. To reduce the false positive probability,
we can break the data frames into a few segments to test the signal level of each segments
separately to ensure that the signal is not from a transient event that has not been properly handled
in the data preparation steps for synthetic tracking velocity search.  We also check whether the
signal is an artifact from the leaking charge or diffraction of a bright object. In practice, these artifacts give most of the false positives. Statistically, for our 12 hours data, we only expect about 0.34\%$ \times 12 \times 3600/30\sim 5$ false detections. However, upon a detection, for confirmation, the false positive rate becomes
\beq
  P_{\rm false-alarm}^{\rm confrim} = \Delta N_x \Delta N_y \Delta N_{v_x} \Delta N_{v_y}
   \times {1 \over 2} {\tt erfc} \Big(
   \frac{\rm SNR }{\sqrt{2}} \Big),
\eeq
where $(\Delta N_x, \Delta N_y)$, and $(\Delta N_{v_x }, \Delta N_{v_y})$ respectively
represent uncertainties in the predicted location and velocity
of the asteroid from the first observation. 

Now the velocity search space only need to cover the uncertainty of the estimated synthetic velocity
$(\Delta N_{v_x }, \Delta N_{v_y})$, which typically does not exceed one grid spacing of the
tracking velocity search grid, which is much smaller than the detection search space of
$N_{v_x} N_{v_y}$ grid points. Therefore, the false positive probability of confirmation
is significantly smaller than that of detection. The uncertainties in determining the location,
on the other hand, grow linearly in time. Therefore, it is important to follow up with the
second observation not too long from the initial detection.
This way the positional uncertainty will be smaller than the FOV.

\section{Results}
\label{sec:results}

\subsection{Detection of a faint object}

We detected a faint asteroid using approximate 30 sec data taken on Sep. 11, 2013 at ${\rm SNR}\sim15$ with tracking velocity 5.97~$\d_p_d$. It was confirmed in the second data set 77 minutes later, which provides another 30 sec of data. The consistency of the velocities and SNRs confirms the detection and also improves the estimation of the
velocity significantly.

\begin{figure}[ht]
\epsscale{0.60}
\includegraphics[width=3.2in, height=3in]{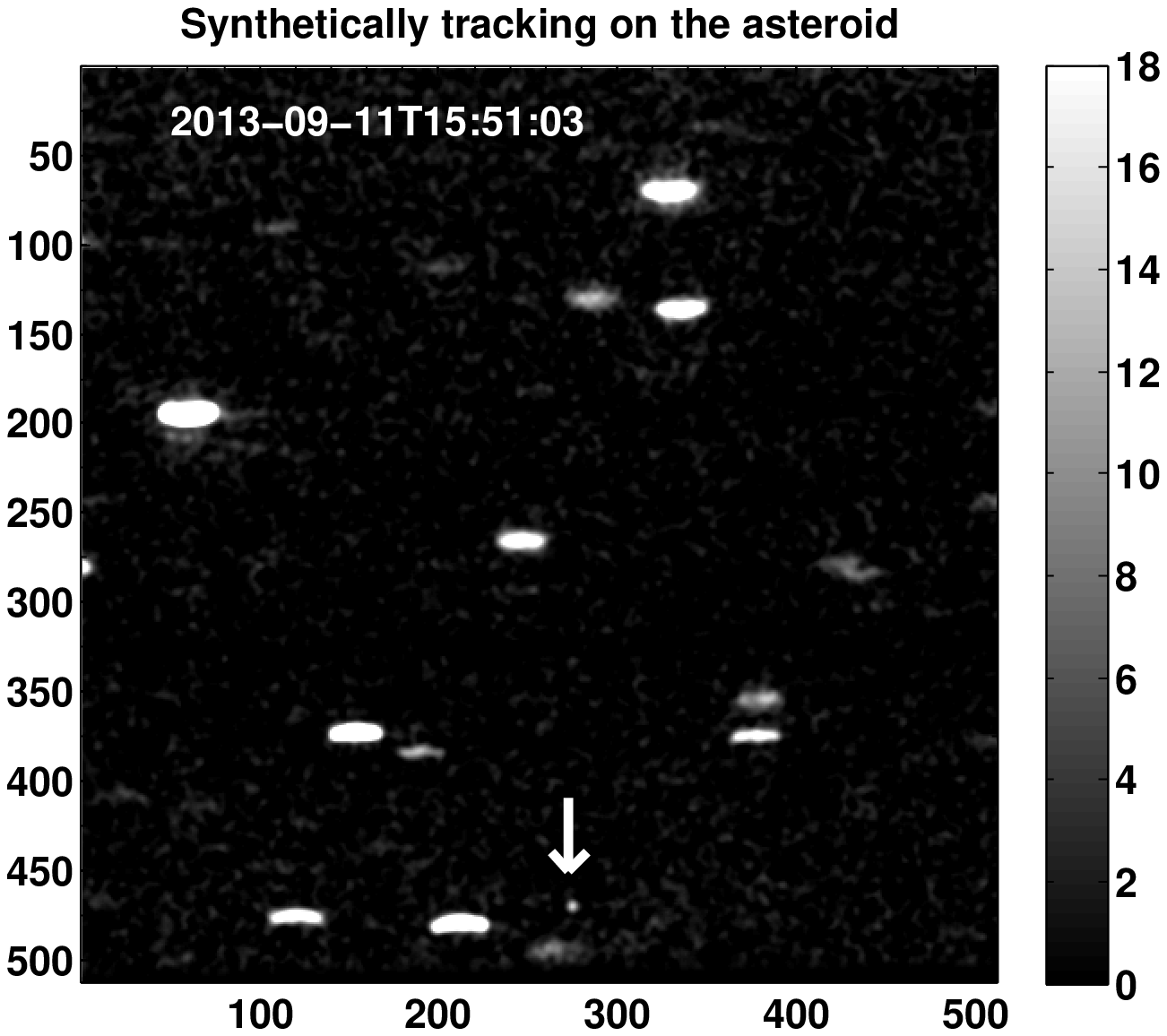}
\includegraphics[width=3.2in, height=3in]{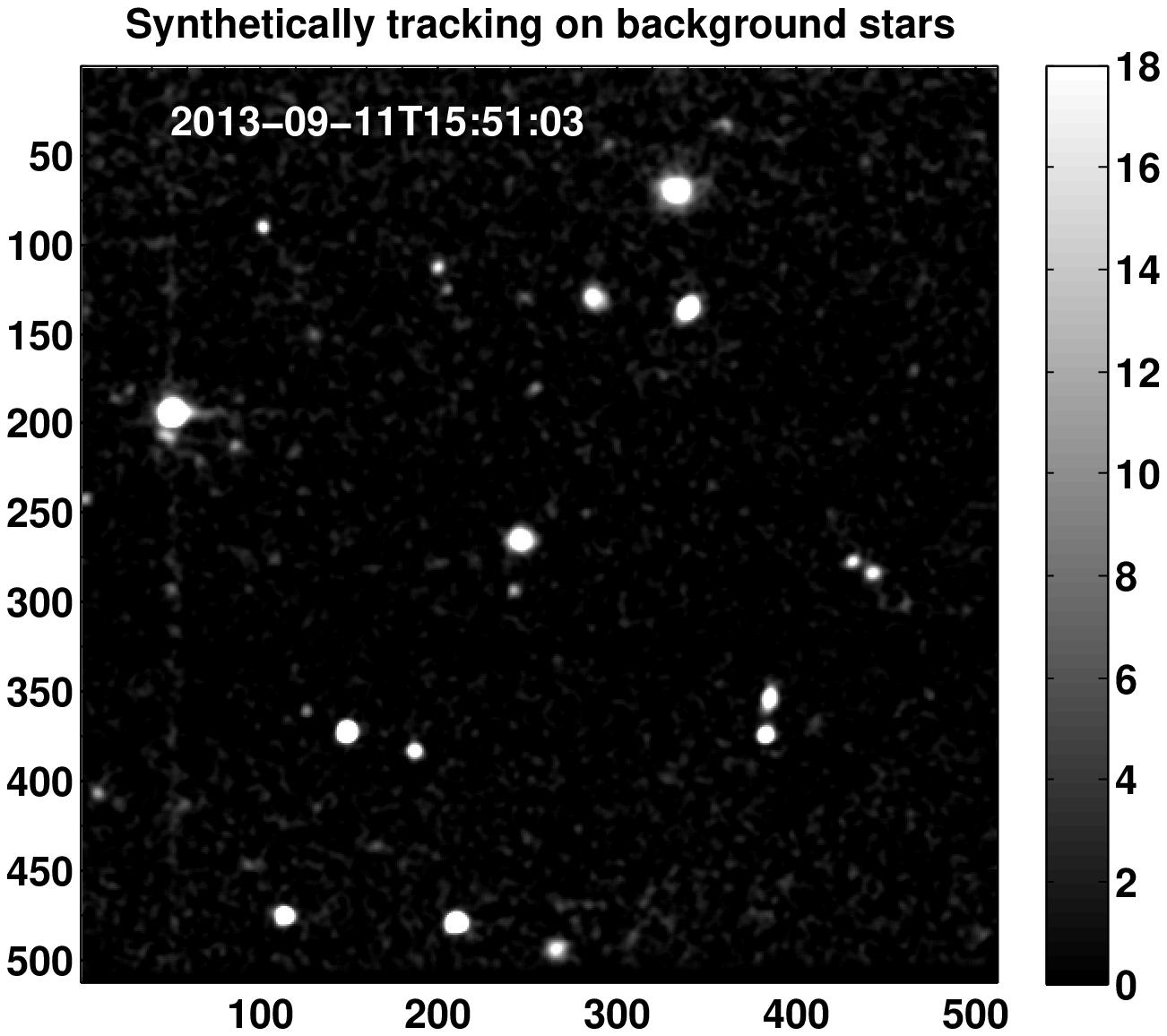}
\caption{Synthetic tracking images on the detected faint asteroid below the arrow (left)
and on the background stars (right), where the faint object is streaked and its surface brightness
is too low to be detected.\label{fig_st_sky_and_nea}}
\end{figure}

Using co-moving PSF fitting, we estimated both the velocities of the asteroid and background stars
with respect to camera frames. Figure~\ref{fig_st_sky_and_nea} shows the IPSA synthetic images
tracking the asteroid and background stars respectively using 524 frames at exposure time 0.06 sec.
The left image, tracking on the asteroid, shows the asteroid at the mid near the bottom as pointed by the arrow.
In the right image, the asteroid can not be identified from the bacground fluctuations due to trailing loss.
The detection SNR is approximately 14.7, giving a false positive probability practically 0 statistically.

\begin{figure}[ht]
\epsscale{0.7}
\includegraphics[width=3.2in, height=3.5in]{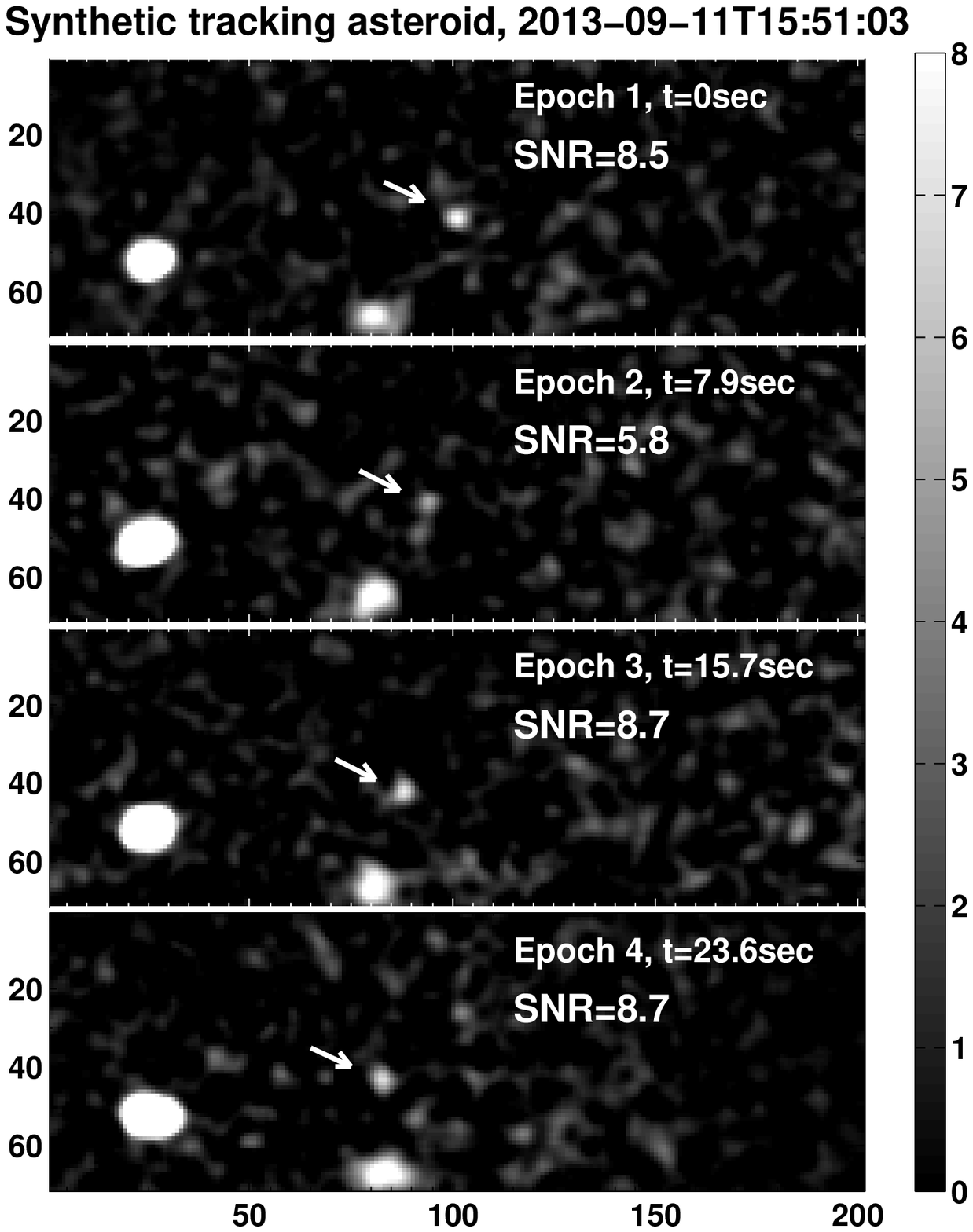}
\includegraphics[width=3.2in, height=3.5in]{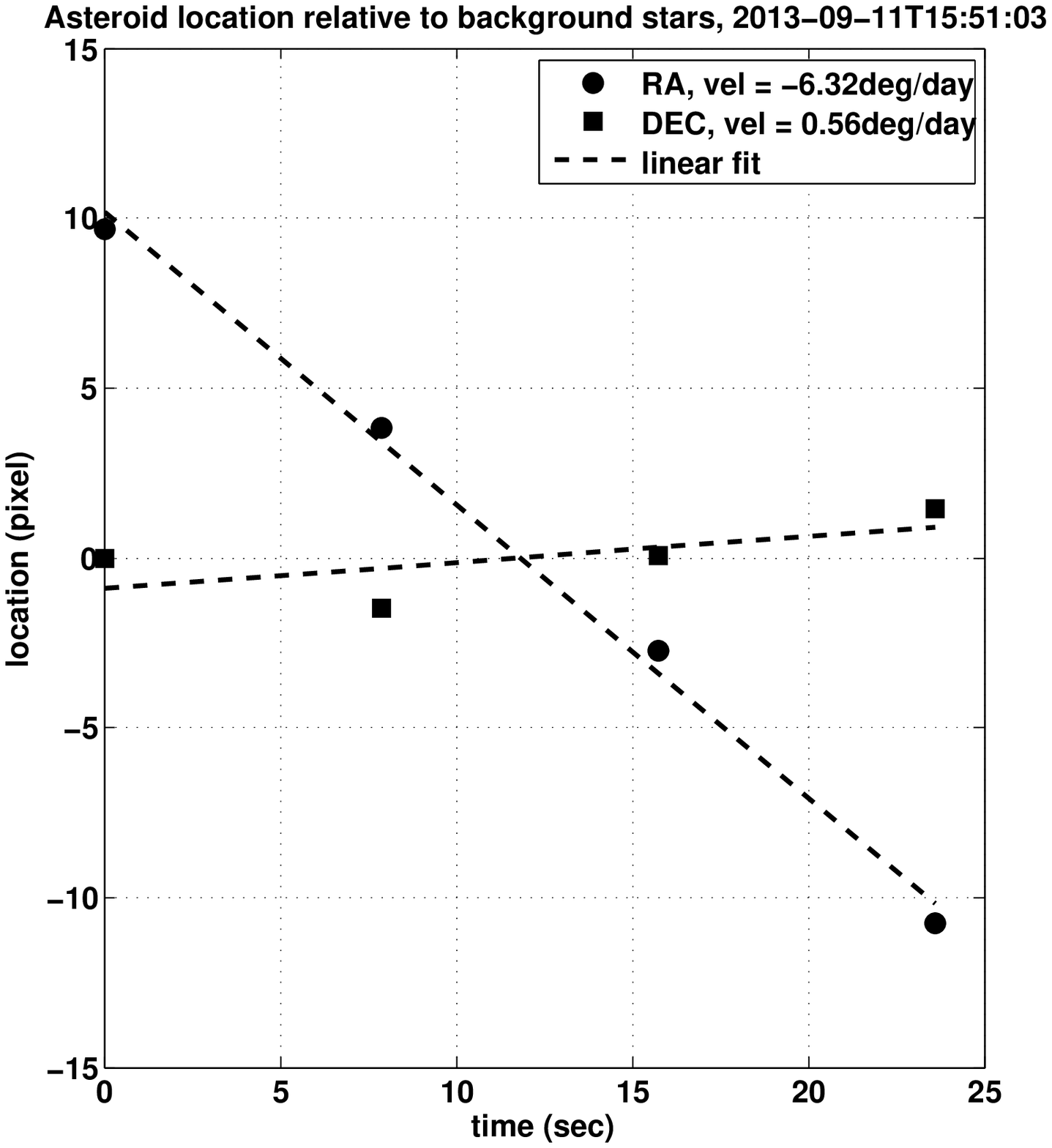}
\caption{(a) Images of the detected asteroid at 4 consecutive epochs separated by 7.8seconds.
The images are genearted by synthetically tracking on the asteroid over the 7.8 seconds.
(b) The relative astrometry of asteroid with respect to the background stars;
an overall constant has been removed from both RA and DEC. \label{fig_4epoch}}
\end{figure}

We broke the data into 4 segments, with each lasting about 7.8 sec (the actual data of observations last
slightly more than 30 seconds), to make sure that the signal shows in all 4 segments of data.
The left four images in Figure~\ref{fig_4epoch} displays the synthetic tracking images of the asteroid at
the four different epochs corresponding to the 4 segments of data. Here we have put the background stars
at approximately the same locations, so that the motion of the asteroid is obvious.
The corresponding SNRs for all the data segments are displayed.

The synthetic tracking yields a trajectory of the object relative to the camera frames.
For the same set of frames, the relative position of objects can be estimated as the difference
between their trajectories. We computed the average of the relative positions between the asteroid and
background stars as the relative astrometry.
The right plot in Fig.~\ref{fig_4epoch} shows the locations of the asteroid relative to the background stars.
The estimated uncertainty of astrometry due to noise is approximately 60 mas for 30 sec observations
using Eq.~(\ref{cent_snr}) (the seeing is approximately 1.2$''$).
The estimated velocity is $[-6.23, 0.04]~\d_p_d$ with precision 0.16~$\d_p_d$ according to Eq.~(\ref{vel_snr}).

\begin{figure}[ht]
\epsscale{1}
\plotone{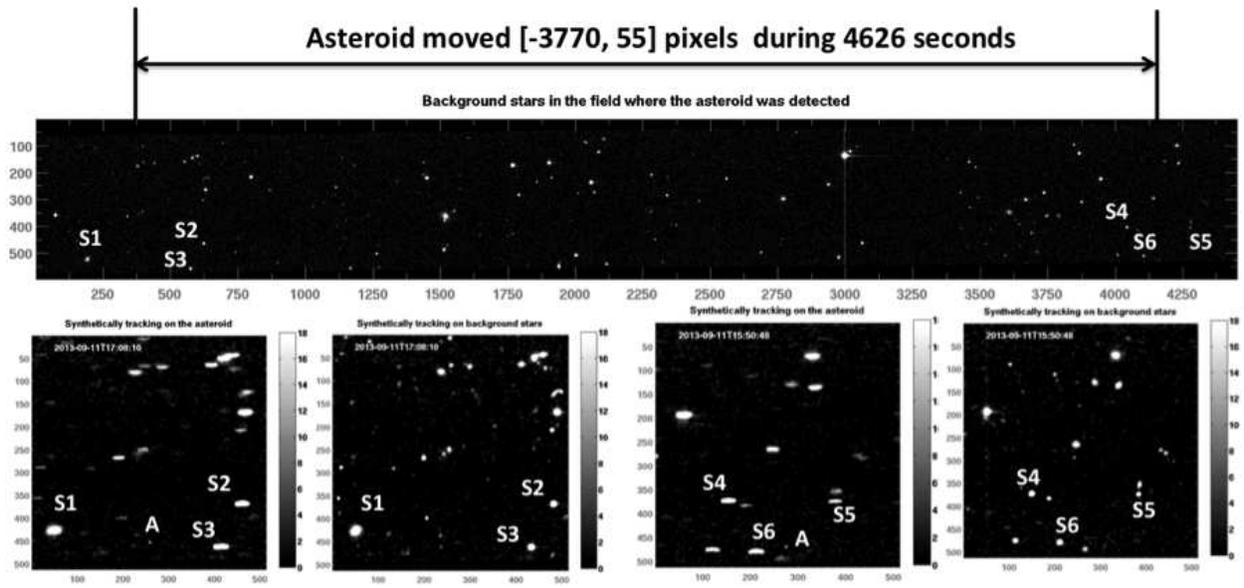}
\caption{Background sky scene (top) covering the range of motion of
the detected asteroid between the two epochs of observations at
2013-09-11T15:50:48 (epoch 1, right two plots at bottom) and 2013-09-11T16:58:24
(epoch 2, left two plots at bottom) respectively. Background stars (S1,S2,S3)
and (S4,S5,S6) are used to compute relative astrometry of asteroid (A) by
synthetically tracking both the asteroid and backgournd stars. From right
to left, the bottom four charts display the IPSA synthetically tracking images
for tracking sky at epoch 1, asteroid at epoch 1, sky at epoch 2, and asteroid
at epoch 2 respectively.\label{fig_sky_two_epochs}}
\end{figure}

The same asteroid was observed 77 mintues later in the second scan over the same field.
The consistency in the velocity and the brightness ensures that the two are the same.
During the two hours observation, the angular distances between stars are assumed to be
constant. As far as there are background stars in the frames where the asteroids are found,
we can estimate its relative motion with respect to those stars and thus the overall
sidereal background as shown in Figure \ref{fig_sky_two_epochs}.
Background stars S4, S5, S6 are used for the first 30 sec observations of the asteroid and
stars S1, S2, and S3 are used for the second observation.
We estimated the relative astrometry of the asteroid with respect to the 
background stars for the two 30 sec observations and then computed the relative positions
between (S1, S2, S3) and (S4, S5, S6) by synthetically tracking at the sidereal rate
(top image in Fig.~\ref{fig_sky_two_epochs}). We found that the asteroid moved totally
$-$3770 pixels along RA and 55 pixel along DEC over 4626.1 seconds,
the separation of two observations, giving velocity of $[-5.97, 0.09]~\d_p_d$
for the faint object on Sep 11, 2013. The accuracy depends
mainly on the accuracy of the plate scale as well as its variation over the field and
the estimation errors in the sidereal rate, which we estimate to be less than 1\%.
The accuracy would be improved dramatically
if we could identify the scene against a star catalog.

The photometry yields approximately 6300 photons over 30 sec integration,
equivalent to a flux of 210~photon/sec. Based on the photometry,
we estimated the apparent magnitude to be 23.1$\pm$0.1 using a system throughput
calibration from observing a known star of magnitude 9.1,
which yielded a flux about 8.1$\times 10^7$~photon/sec at the detector.
As we did before, we divided the total data from the two 30-sec observations into two 4-segments
with each lasting about 7.8 sec and computed the standard deviations of photometry over these 8 segments
to estimated the 0.1 magnitude uncertainty in the photometry.

Assuming a typical asteroid with speed of 10 km/s relative to the Earth, we estimate the distance of the object is approximately at a distance of $d\sim$ (10 km/s)/(5.97~$\d_p_d$) $\sim$ 8$\times 10^6$ km $\sim$ 20 lunar
distances. Assuming albedo of 0.15, we estimate the size of the asteroid to be 8 m.

The official discovery requires observing asteroid
at a minimal three epochs to determine its orbit.
Even though the false positive probability is almost zero,
we cannot claim discovery of the asteroid because
we only observed the asteroid at two epochs on the same night.

\subsection{Astrometry precision of two know asteroids}
\label{sec:astrometry}

On Apr 3, 2013, we observed two known asteroids 2009BL$_2$ and 2013FQ$_{10}$. We compare the astrometry obtained by using synthetic tracking technique and traditional long exposure streaked images, simulated by co-adding short exposure images, to show the improvement of precision from the improved SNR and cancellation of effects due to atmosphere and imprecise telescope pointing in relative astrometry.

\begin{figure}[ht]
\epsscale{0.6}
\plotone{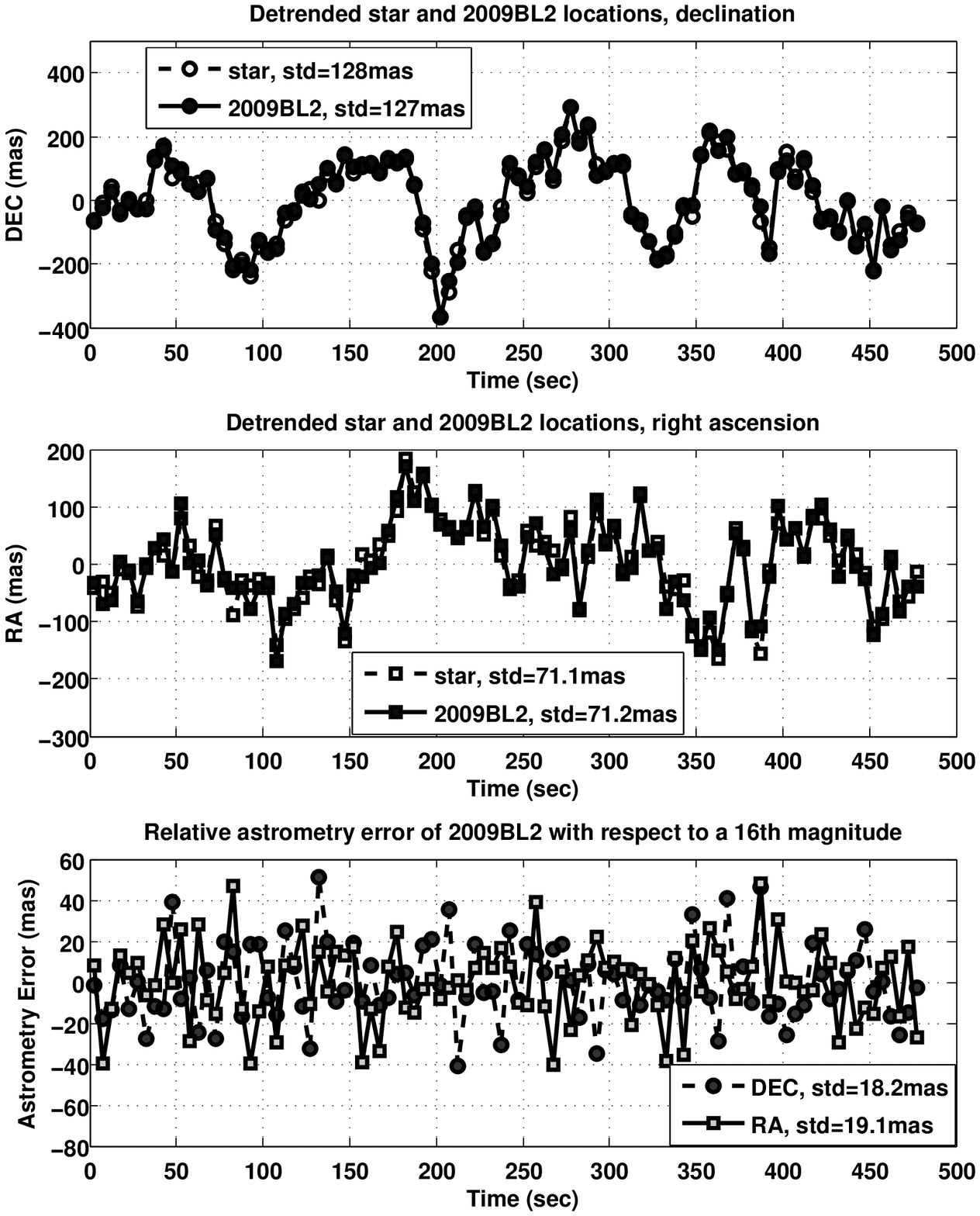}
\caption{Temporal variations of the locations of the 16th magintude background star
and the asteroid 2009BL$_2$ with respect to the camera frame (top for DEC and mid for
RA) and their differences (bottom, relative astrometry). They are detrended
by removing an overall constant plus a linear trend. The top two panels show 100~mas
level of variation of astrometry, caused by atmosphere and imprecising pointing of
the telescope, are common to both the star and asteroid. The small RMS in the bottom 
panel shows the cancellation of this common variation.
\label{fig_star_2009BL2_detrend}}
\end{figure}

Because asteroid moves at approximately constant velocity during the 15 minutes of observation, we compute the relative astrometry between
the asteroid and background stars and compare it with a constant
motion to determine errors in the astrometry. The top and mid plots
in Figure \ref{fig_star_2009BL2_detrend} show the
de-trended temporal variation of the location of a 16th magnitude background star (empty dots/squares) and the location of asteroid 2009BL$_2$ (solid dots/squares) (of apparent magnitude 18.5) relative to the camera frames
(telescope pointing), for declination (top) and right ascension (mid) respectively. In the field, the star locations vary due to atmosphere effect and imprecise
telescope pointing with an RMS of 100 mas. We de-trended (by removing an overall constant and a linear trend from the data) both the star locations (there is a small drift in the tracking) and asteroid locations to see the common temporal variation of the locations. The bottom plot in Figure \ref{fig_star_2009BL2_detrend} displays the de-trended relative astrometry of the asteroid with respect to the 16th magnitude star, which is mostly linear (not shown, de-trended to see the residual) with less
than 20 mas residual RMS, much smaller than 100 mas level variation. This shows the cancellation of of the common temporal variations shown in top and mid plots between asteroid and star. However, this does not happen if we co-add the images to synthesize a long exposure.

\begin{figure}[ht]
\epsscale{0.6}
\plotone{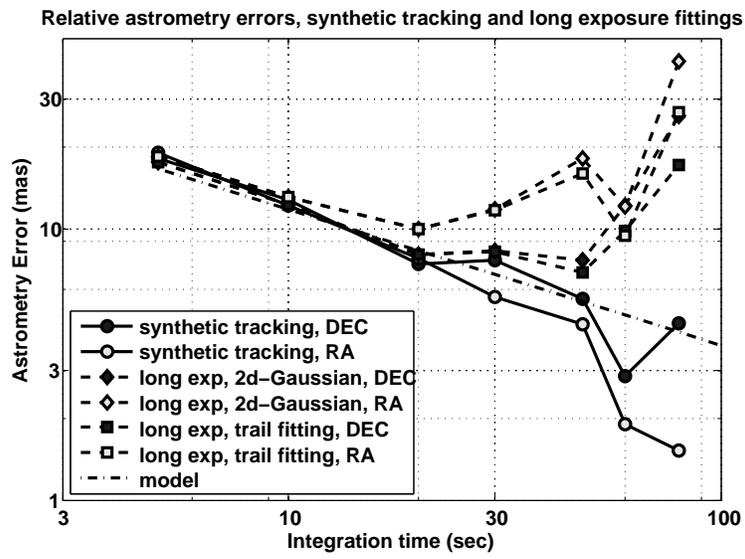}
\caption{Aastrometry error RMS as function of integration time. Three cases:
synthetic tracking (circle marker), traditional long exposure fitting a 2-d Gaussian PSF
(diamond marker), traditional long exposure fitting a Gaussian trail (square marker),
are compared.\label{fig_err_RMS_vs_exp}}
\end{figure}

Figure \ref{fig_err_RMS_vs_exp} shows the astrometry error RMS as function of the integration time.
The three pairs of curves show the RMSs of the astrometry errors
for using synthetic tracking with short exposures (dots) and
a single exposure with the duration being the integration time and processed
using a conventional centroid fitting to a 2-d Gaussian PSF (diamond) or a Gaussian trail function (squares)
as described in \citep{Veres2012}.
The solid and empty markers denote declination and right ascension respectively.
We can see the performance difference between the synthetic tracking and the conventional long exposure results.
Comparing with the estimated photon noise limited error RMSs (dot dashed line) using Eq.~(\ref{comoving_lsq}),
as the integration time increases, the synthetic tracking astrometry error is close to be photon noise limited
and decreases as the inverse of the square root of the integration time.
However, the traditional long exposure approach does not improve as the integration
time goes beyond 30 sec because the image of the asteroid is streaked by more than 1$''$ and
the error is not dominated by the photon noise, but by the effects from atmospheric
and imprecise telescope pointing, which is no longer common between the streaked asteroid and the background star images.

\begin{figure}[ht]
\epsscale{0.6}
\plotone{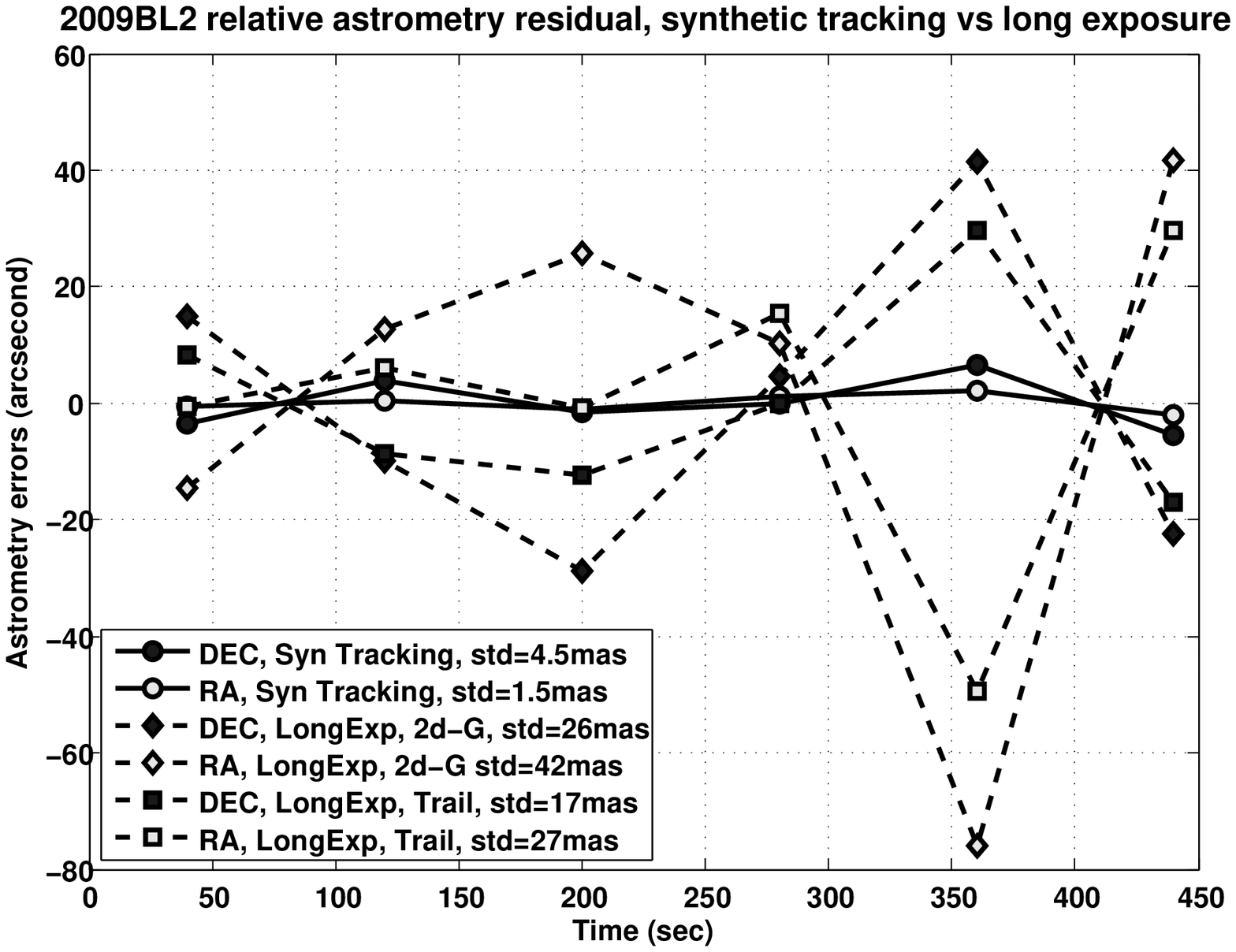}
\caption{Residuals from fitting the asteroid astrometry relative to the
16th magntiude star to a linear motion model. Each data point is obtained
from integrating 80 seconds of data. Three sets of curves are respectively for
synthetic tracking (solid curves with circle markers), single exposure using 2-d Gaussian
centroiding (dashed curves with diamond markers), and single exposure using Gaussian trail
centroiding (dashed curves with square markers).\label{fig_rel_astr_res_cmp}}
\end{figure}

Figure \ref{fig_rel_astr_res_cmp} displays the residual of relative astrometry,
after fitting to a linear motion, with 80 sec integration time, for the cases of using synthetic
tracking (circles), long exposure with a 2-d Gaussian fitting
(diamonds), and long exposure with a Gaussian trail fitting (squares) respectively.
The synthetic tracking using short exposures enables to achieve mas-level astrometry
precision, close to being photon and background noise limited as in relative stellar astrometry \citep{Boss2009}.
Using long exposure, on the other hand, the errors due to atmosphere and imprecise telescope
pointing are non-common between the asteroid and stars and thus lead to tens of milliarcsecond errors
in relative astrometry. These are results of using telescope of 5 m.
The atmospheric effect for using 1-m class telescopes is expected to be even larger.

\begin{figure}[ht]
\epsscale{0.6}
\plotone{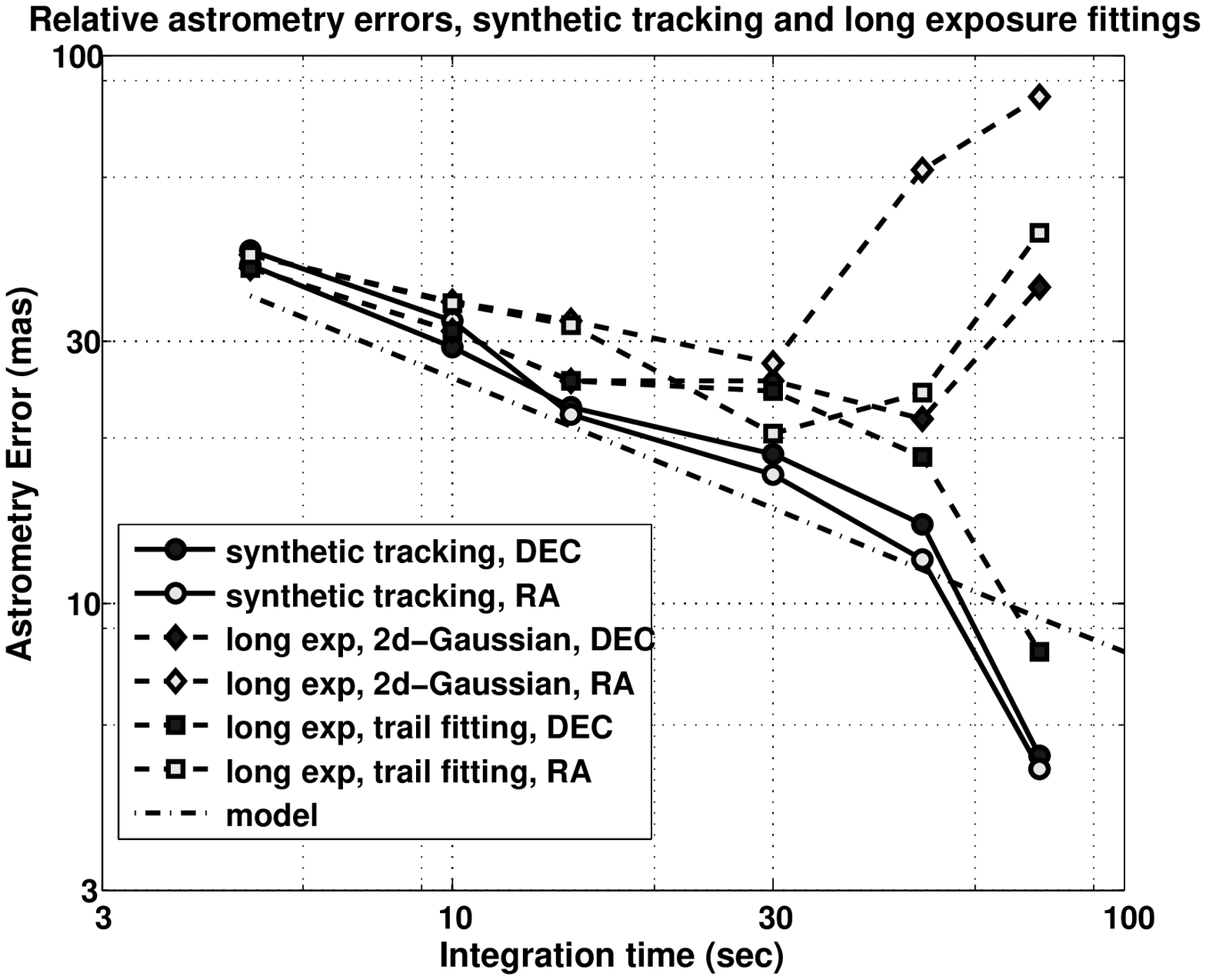}
\caption{Relative astrometric error of asteroid 2013FQ$_{10}$ relative to a 19th magnitude star in the field.
Three sets of curves are respectively for synthetic tracking
(solid curves with circle markers), single exposure using 2-d Gaussian
centroiding (dashed curves with diamond markers), and single exposure using Gaussian trail
centroiding (dashed curves with square markers).\label{fig_2013FQ10_RMS}}
\end{figure}

We obtained similar results for the asteroid 2013FQ$_{10}$,
which we observed for 300~sec. Figure~\ref{fig_2013FQ10_RMS}, 
which is the same plot as Fig.~\ref{fig_err_RMS_vs_exp} for 2013FQ$_{10}$,
shows the improvement of astrometry as we integrate more 2 Hz frames.
The residual errors are mainly due to photon and background noise as estimated using
the empirical formula Eq.~(\ref{cent_snr}).
The error becomes single digit of mas after integrating over 100~sec.
Again, if we would use long exposure ({\it i.e}., 30 sec) images,
even with the state of art fitting of a Gaussian trail (square markers),
the error would still be as large as tens of mas, growing with the integration time.

\subsection{Discovery and orbit determination of NEAs with synthetic tracking}

Synthetic tracking is especially valuable for detecting very small and fast moving asteroids. The vast majorities of these objects are so small and move so rapidly that follow-up observations have to be planned from the start. Finding the object one hour later when it moved 1000$''$ is possible only
because our single observation has a coarse velocity measurement.
For really faint object with ${\rm SNR} \sim 7$, the coarse velocity measurement with uncertainty
$\sim$ 0.3~$\d_p_d$, would not let another observatory find it one day later easily. In the recent observing run, we scanned each part of the sky twice. However, a large portion of that time is wasted. Ideally we would have the GPU software running to detect the NEAs in near real time so that the second confirming observation could be made
within the time frame (a few hours) later before the uncertainty in the location of the
asteroid becomes larger than the FOV.
Once confirmed we should spend significantly more than 30 sec to get astrometry approaching
50~mas so that three observations
spaced a few days apart would let us derive an orbit where the object could be observed again at the next apparition \citep{Giorgini2013}. Objects with H$\sim$28--30 moving at 6~$\d_p_d$ are detectable only on medium to large telescopes with synthetic tracking cameras, and even then, only for a week or so before they are too faint to be detected.  Therefore, it is crucial to have a strategy like discussed above to discover the object within its time window for observation.

\section{Conclusions}
\label{sec:conclude}

We have presented the detection of a faint object as an application of the synthetic tracking technique to observational data taken on the Palomar 200-inch telescope including the data processing method and algorithms to demonstrate the efficacy. Synthetic tracking
significantly improved detection SNR over traditional long exposure approach for fast moving NEAs, and thus enabled the detection of this faint object at apparent magnitude of 23. It also yields more precise astrometry by improving SNR and making the effects due to atmosphere and imprecise telescope pointing common between the asteroid and background stars, which cancels to the first order in relatively astrometry. Using the observational data of two known asteroids, we demonstrated milli aresecond level
precision in astrometry. From a 12-hour blind search we found only one asteroid, which is consistent with Harris's population distribution of asteroids within the order of magnitude. Observation strategy using synthetic tracking to detect and discover faint objects within the short observation time window in general requires a well scheduled
observation including a close to real time detection using 30 sec of observation data, confirmation within a few hours, and orbit determination within a week before the object becomes too faint to observe.

\acknowledgments
The authors would like to thank Dr.~Stuart Shaklan and Dr.~William Owen of JPL for useful discussions, and Dr.~Eric Cady for editing the abstract.
The work described here was carried out at the Jet Propulsion Laboratory, California Institute of Technology,
under a contract with the National Aeronautics and Space Administration.
Copyright 2014. Government sponsorship acknowledged.

\clearpage




\clearpage








\clearpage


\clearpage




\end{document}